\def\mode{0}
\if 1\mode
	\documentclass[10pt]{wlscirep}
\else
	\documentclass{article}
\fi

\usepackage{etoolbox}
\newtoggle{scirep}
\if 1\mode
	\toggletrue{scirep}
\else
	\togglefalse{scirep}
\fi

\iftoggle{scirep}{
	\usepackage[nomarkers, nolists]{endfloat}
}{
	\usepackage[
	a4paper,
	left=25mm,
	right=25mm,
	top=20mm,
	bottom=34mm,
	]{geometry}
	\usepackage{amsmath}
	\usepackage{authblk}
	\usepackage{url}
	\usepackage{hyperref}
	\bibliographystyle{mrm}
	\usepackage{graphicx}

	\newcommand{\keywords}[1]{\textbf{Key words: #1}}
}

\usepackage[utf8]{inputenc}

\usepackage{bm}
\DeclareMathAlphabet\mathcal{OMS}{cmsy}{m}{n}
\SetMathAlphabet\mathcal{bold}{OMS}{cmsy}{b}{n}

\usepackage[
	capitalise,
]{cleveref}

\usepackage{changes}
\newcommand{\stkout}[1]{\ifmmode\text{\sout{\ensuremath{#1}}}\else\sout{#1}\fi}
\setdeletedmarkup{\stkout{#1}}

\usepackage{ifdraft}
\usepackage{marginnote}

\usepackage{xfrac} 

\DeclareMathOperator*{\argmin}{arg\,min}
\def\defd{\mathrel{\mathop:}=}
\def\dfed{=\mathrel{\mathop:}}
\newcommand*{\vect}[1]{\bm{#1}}

\usepackage[
per-mode=fraction,
separate-uncertainty,
retain-unity-mantissa=false,
product-units=power,
table-alignment=center,
detect-all,
binary-units=true,
exponent-product=\cdot,
exponent-base=10
]{siunitx}

\title{ENLIVE: An Efficient Nonlinear Method for Calibrationless and Robust Parallel Imaging}
\author[1,2,*]{H. Christian M. Holme}
\author[1,2]{Sebastian Rosenzweig}
\author[3]{Frank Ong}
\author[1,2]{Robin N. Wilke}
\author[3]{Michael Lustig}
\author[1,2]{Martin Uecker}
\affil[1]{Institute for Diagnostic and Interventional Radiology,
	University Medical Center Göttingen, Göttingen, Germany}
\affil[2]{German Centre for Cardiovascular Research (DZHK),
	Partner site Göttingen, Göttingen, Germany}
\affil[3]{Departement of Electrical Engineering and Computer Sciences,
	University of California, Berkeley}
\affil[*]{\texttt{christian.holme@med.uni-goettingen.de}}

\def\myabstract{
\begin{abstract}
	Robustness against data inconsistencies, imaging artifacts and acquisition
	speed are crucial factors limiting the possible
	range of applications for magnetic resonance imaging (MRI). Therefore, we
	report a novel calibrationless parallel
	imaging technique which simultaneously estimates coil profiles and image
	content in a relaxed forward model. Our
	method is robust against a wide class of data inconsistencies, minimizes
	imaging artifacts and is comparably fast
	combining important advantages of many conceptually different
	state-of-the-art parallel imaging approaches.
	Depending on the experimental setting, data can be undersampled well below
	the Nyquist limit.
	Here, even high acceleration factors yield excellent imaging results while
	being robust to noise and the occurrence of phase
	singularities in the image domain, as we show on different data.
	Moreover, our method successfully reconstructs acquisitions with
	insufficient field-of-view. We further compare our approach to ESPIRiT
	and SAKE using spin-echo and gradient echo MRI data from the human head
	and knee.
	In addition,  we show its applicability to non-Cartesian imaging on radial 
	FLASH cardiac MRI data.
	Using theoretical considerations, we show that ENLIVE can be related
	to a low-rank formulation of blind multi-channel deconvolution, explaining
	why it inherently promotes low-rank solutions.
\end{abstract}
}

\def\mykeywords{
\keywords{magnetic resonance imaging, parallel imaging, \mbox{ESPIRiT},
iterative reconstruction, low-rank constraint, NLINV}
}

\iftoggle{scirep}{
\myabstract
\mykeywords

\begin{document}

	\flushbottom
	\maketitle
}{
\newlength{\savedparindent}
\setlength{\savedparindent}{\parindent}

\begin{document}

	\setlength{\parindent}{0in}
	\flushbottom
	\maketitle

	\myabstract

	\vspace{2cm}

	\mykeywords

	\newpage

	\setlength{\parindent}{\savedparindent}
}

\section*{Introduction}
Since acquisition speed is a major issue in MRI, accelerated imaging
with multiple receiver coils has been an active field of research since its
beginning. Quite rapidly, two main categories
of parallel imaging methods emerged: image-space methods, of
which sensitivity encoding (SENSE)~\cite{Pruessmann_Magn.Reson.Med._1999} is
the prototypical example and
k-space methods, where it is generalized autocalibrating partially parallel
ac\-qui\-si\-tions (GRAPPA)~\cite{Griswold_Magn.Reson.Med._2002}. SENSE-like 
methods,
when the coil sensitivity profiles are known, permit a natural description as
a linear inverse problem. Incorporating the estimation
of coil sensitivity profiles into the reconstruction
leads to a nonlinear inverse problem, as formulated in
Joint Image Reconstruction and Sensitivity Estimation in SENSE
(JSENSE)~\cite{Ying_Magn.Reson.Med._2007}
and Regularized Nonlinear Inversion
(NLINV)~\cite{Uecker_Magn.Reson.Med._2008}.

Additionally, low-rank and subspace 
methods~\cite{Shin_Magn.Reson.Med._2014,
Haldar_IEEETrans.Med.Imaging_2014,
Haldar_Magn.Reson.Med._2016,
Trzasko__2011}
have been proposed to further increase
reliability and acceleration in MRI. These methods exploit prior knowledge
on the structure of the matrices arising in MRI reconstruction. Recently,
ESPIRiT~\cite{Uecker_Magn.Reson.Med._2014} has been shown to
provide robustness towards data
inconsistencies similar to k-space methods such as GRAPPA~\cite{Griswold_Magn.Reson.Med._2002}.
In particular, in cases where the chosen field-of-view (FOV)
is smaller than the object~\cite{Griswold_Magn.Reson.Med._2004} and in
phase-constraint imaging~\cite{Uecker_Magn.Reson.Med._2017}, it was
shown that methods based on traditional SENSE
that only use a single set of coil sensitivity profiles exhibit artifacts.
In ESPIRiT, robust reconstruction is possible through a relaxed
SENSE-model, which uses multiple images and sets of coil sensitivity profiles.

ESPIRiT recovers accurate coil sensitivities using an eigenvalue decomposition
of an image-domain operator which projects onto the signal space
of the calibration matrix. In case of inconsistencies, it produces multiple sets
of maps which can be used in a relaxed SENSE reconstruction.
ESPIRiT requires a fully-sampled calibration region in the center of k-space.
Additionally, it cannot be applied directly to non-Cartesian data, requiring
an additional gridding step to generate calibration data.
A more generic subspace method is SAKE~\cite{Shin_Magn.Reson.Med._2014},
because it can be directly applied
to data without fully-sampled calibration region or non-Cartesian data.
Based on the idea that the signal
is contained in a sub-space of smaller dimensionality which can be recovered,
SAKE uses structured low-rank matrix 
completion to recover a full k-space from incomplete data. 
Unfortunately, it is computationally extremely demanding as each
iteration has to perform a singular-value decomposition (SVD). Furthermore,
because it operates completely in k-space, regularization
terms may require additional Fourier transforms and must be applied 
to all channels.
Calibration-free locally low-rank encouraging 
reconstruction (CLEAR)~\cite{Trzasko__2011} is a related method which uses
local low-rankness in the image domain instead of the global k-space rank 
penalty used in SAKE. This reduces the computational complexity by reducing the
size of the needed SVDs, although it does increase the number of SVDs
necessary. Furthermore, as it is an image space method,
regularization can be integrated more easily.

Regularized Nonlinear Inversion 
(NLINV)~\cite{Uecker_Magn.Reson.Med._2008} jointly estimates the
image content and the coil
sensitivity profiles using a nonlinear algorithm. Similar to SAKE, it
does not require a fully-sampled Cartesian calibration region and
can be applied directly to non-Cartesian data.

This work aims at combining the advantages from these different methods.
Inspired by ESPIRiT, we propose an extension to NLINV that extends
it beyond the original SENSE-like model.
This method, termed ENLIVE (Extended NonLinear InVersion inspired by ESPIRiT),
can be  related to a convex relaxation of the NLINV problem
subject to a low-rank constraint.
From NLINV, it inherits its flexibility and suitability for calibrationless
and non-Cartesian imaging;
from ESPIRiT it inherits robustness to data inconsistencies. We apply ENLIVE
to several imaging settings covering limited FOV, phase constraints,
phase singularities, and non-Cartesian acquisition.
Additionally, we present comparisons to ESPIRiT and
to SAKE.

Initial results have been presented at the
25\textsuperscript{th} Annual Meeting of the International Society for Magnetic
Resonance in Medicine.\cite{Holme__2017}

\section*{Theory}

\subsection*{Formulation}

NLINV recovers the image $\vect{m}$ and the coil sensitivity profiles 
$\vect{c}_j$
from measurements $\vect{y}_j$ by solving
the regularized nonlinear optimization problem:
\begin{align}
\argmin_{\vect{m}, \vect{c}_j}\: \sum_{j=1}^{N_{C}} \| \vect{y}_j - 
\mathcal{P}\mathcal{F} \{
\vect{c}_j\odot \vect{m} \}\|_2^2 + \alpha ( \sum_{j=1}^{N_{C}}\| \vect{W} 
\vect{c}_j \|_2^2
+ \|\vect{m}\|_2^2 )
\label{eq:nlinv}
\end{align}
with $N_{C}$ coils, the
two or three dimensional
Fourier transform $\mathcal{F}$, the 
projection 
$\mathcal{P}$ onto
the measured trajectory (or the acquired pattern in Cartesian imaging)
and an invertible weighting matrix $\vect{W}$ penalizing high frequencies in 
the coil
profiles.
Here, both image $\vect{m}\in \mathcal{C}^{n_x\cdot n_y\cdot n_z}$ and
coils $\vect{c}_j\in \mathcal{C}^{n_x\cdot n_y\cdot n_z}$ are regarded as 
vectors
of size $n_x\cdot n_y\cdot n_z\dfed N_{I}$ and
$\odot$ is their element-wise product.

In this work, we propose to extend this model to:
\begin{align}
\argmin_{\vect{m}^i,\vect{c}_j^i}\: \sum_{j=1}^{N_{C}}\|\vect{y}_j -
\mathcal{PF}\{\sum_{i=1}^k\vect{c}^i_j\odot \vect{m}^i\}\|^2_2 +
\alpha
\sum_{i=1}^{k}
(\sum_{j=1}^{N_{C}}\|\vect{W}\vect{c}^i_j\|_2^2 +
\|\vect{m}^i\|_2^2)
\label{eq:enlive}
\end{align}
where $\vect{c}_j^i$ and $\vect{m}^i$ are $k$ sets
of unknown coil sensitivity profiles and unknown images.
This approach is inspired by ESPIRiT, which uses additional
maps to account for model violations.\cite{Uecker_Magn.Reson.Med._2014}

In the following, we will show that this formulation automatically
produces solutions with rank even smaller than $k$ if one exits.
To show this, we first relate \cref{eq:enlive} to a linear inverse
problem for matrices with nuclear norm regularization.

From here on, we assume that the variable transformation
$\hat{\vect{c}}_j = \vect{W} \vect{c}_j$ has been applied to move the weighting 
matrix from
the regularization into the forward operator.
We note that this problem is equivalent to a corresponding
multi-channel blind deconvolution
problem~\cite{Kundur_IEEESignalProcess.Mag._1996} in k-space via
the convolution theorem.
Using the "lifting" approach used for such blind deconvolution
problems~\cite{Ahmed_IEEETrans.Inform.Theory_2014}, which can also be applied
in the image domain, we now lift the \cref{eq:nlinv}
into a linear
inverse problem in terms of a rank-1 matrix $\vect{X} = \vect{u}\vect{v}^T$
formed by the tensor product of $\vect{u}$ and $\vect{v}$,
where $\vect{u}$ corresponds to $\vect{m}$ and $\vect{v}$ is a stacked vector 
composed of the
weighted coil
sensitivity profiles $\hat{\vect{c}}_j$. The problem then becomes:
\begin{align}
\argmin_{\vect{u},\vect{v}}\: \| \vect{y} - \mathcal{A}\{\vect{u}\vect{v}^T\} 
\|_2^2 + \alpha ( 
\| \vect{u} \|_2^2 + \|\vect{v}\|_2^2 )
\label{eq:outerp_nlinv}
\end{align}
with a linear operator $\mathcal{A}$ mapping $\vect{u}\vect{v}^T$ to 
$\mathcal{PF}\vect{c}_j\odot \vect{m}$
and a vector $\vect{y}$ containing measurement data of all coils. Such
an $\mathcal{A}$ exists because $\vect{u}\vect{v}^T$ contains all possible
products of elements of $\vect{u}$ and $\vect{v}$.
Its explicit action is explained in more detail in the Appendix.
In general, all bilinear
functions can be expressed as linear functions on the
tensor product of the two vector spaces involved.

As suggested by Ahmed et al.~\cite{Ahmed_IEEETrans.Inform.Theory_2014} 
for blind multi-channel deconvolution,
we now relax
the rank-1 constraint and allow $k$ sets of images and
coil sensitivity profiles.
This corresponds
to using $\vect{X}=\vect{U}\vect{V}^T\in
\mathcal{C}^{N_{I}\times N_{C}\cdot N_{I}}$
with $\vect{U}\in \mathcal{\vect{C}}^{N_{I}\times k}$
and $\vect{V}\in \mathcal{C}^{N_{C}\cdot N_{I}\times k}$, which then leads
to the optimization problem
\begin{align}
\argmin_{\vect{U},\vect{V}}\: \| \vect{y} - \mathcal{A}\{\vect{U}\vect{V}^T\} 
\|_2^2 + \alpha ( 
\| \vect{U} \|_F^2 + \|\vect{V}\|_F^2 )
\label{eq:lifted_nlinv}
\end{align}
with the Frobenius norm $\|\cdot\|_F$.
In the Appendix we show how this corresponds to ENLIVE as
formulated in \Cref{eq:enlive}.
Under conditions given below, \cref{eq:lifted_nlinv} is equivalent
to a convex optimization problem
for the matrix
\begin{align}
\argmin_{\vect{X}}\: \| \vect{y} - \mathcal{A}\{\vect{X}\} \|_2^2 + 2 \alpha 
\|\vect{X}\|_\star
\label{eq:convex_relax}
\end{align}
with nuclear norm $\|\cdot\|_\star$
regularization.\cite{Davenport_IeeeJ.Sel.Top.Signa._2016,Recht_SIAMReview_2010}
The nuclear norm promotes low-rank solutions. Furthermore, if the
solution to \cref{eq:convex_relax} has rank smaller than or equal to $k$ 
both problems are equivalent in the sense that from a 
solution $\vect{U},\vect{V}$ of \Cref{eq:lifted_nlinv} one obtains a solution
of \Cref{eq:convex_relax} via $\vect{X} = \vect{U} \vect{V}^T$ which attains the
same value and from a solution $\vect{X}$ of \Cref{eq:convex_relax}
one can construct a solution of \Cref{eq:lifted_nlinv} that
attains the same value. This is achieved by factorizing $\vect{X}$
using the SVD and by distributing the singular values 
in an optimal way, i.e. equally as square roots, to the two 
factors.
Please note that we do not propose to use this convex formulation for
computation as it is very expensive, instead we propose to use
the nonlinear
formulation given in \cref{eq:enlive}. Nevertheless, this relationship 
to nuclear-norm regularization is important as it explains why ENLIVE produces 
solutions with low
rank even smaller than $k$, if one exists.

\subsection*{Implementation}

Similar to NLINV~\cite{Uecker_Magn.Reson.Med._2008}, we solve
\cref{eq:enlive} using the iteratively regularized Gauss-Newton method (IRGNM).
The IRGNM solves successive linearizations
with the regularization parameter decreasing in each Newton step:
Starting from $\alpha_0$,
the regularizations in each step is reduced according to
$\alpha_n=\alpha_0q^{n-1},\,0<q<1$.
As initial guess, we use
$\vect{m}^i \equiv 1$ for the images and $\vect{c}^i_j\equiv 0$ for the coil 
sensitivity
profiles. Because we initialize images and  sensitivity profiles for all sets
in the same way, the problem is symmetric in the sets and the algorithm will
produce degenerate solutions with identical sets.
To break this symmetry, we require the $k$ sets of coil profiles to be
orthogonal using Gram-Schmidt orthogonalization after each Newton step.
For orthogonalization, the coil profiles of each set are treated as
stacked one-dimensional vectors.

The weighting matrix $\vect{W}$ enforcing smoothness in the coil
profiles was chosen as in \cite{Uecker_Magn.Reson.Med._2008}. In k-space, this 
leads to a penalty increasing with distance from the center of k-space
according to $(1+a\|\vect{k}\|^2)^{b/2}$. In this work, $a=240$ and $b=40$ were
used. Furthermore, k-space is normalized so that it extends from 
$\sfrac{-n_i}{2}$ to $\sfrac{n_i}{2}$ for $i\in\{x,y,z\}$. As $\vect{W}$ applies
weights in k-space, it is the product of a Fourier matrix transforming each
coil profile to k-space an of this diagonal weight matrix.

Images and coil profiles are combined in a post-processing step. This is
used to either create individual images for each set~$i$ by
combining coil-weighted images $\vect{m}^i \vect{c}_j^i$ using
\begin{align}
	\vect{M}^i = \sqrt{\sum_{j=1}^{N_{C}}|\vect{m^}i\odot \vect{c}^i_j|^2}
\end{align}
or to create a single combined image by first combining each
set to obtain a proper image for each coil and then doing a
final coil combination with
\begin{align}
\vect{M} = 
\sqrt{\sum_{j=1}^{N_{C}}\bigg|\sum_{i=1}^{k}\vect{m}^i\odot 
\vect{c}^i_j\bigg|^2}.
\end{align}

\section*{Results}

\newcommand{\figscale}{0.35}
\begin{figure}[htbp]
	\centering
	\includegraphics[scale=\figscale]{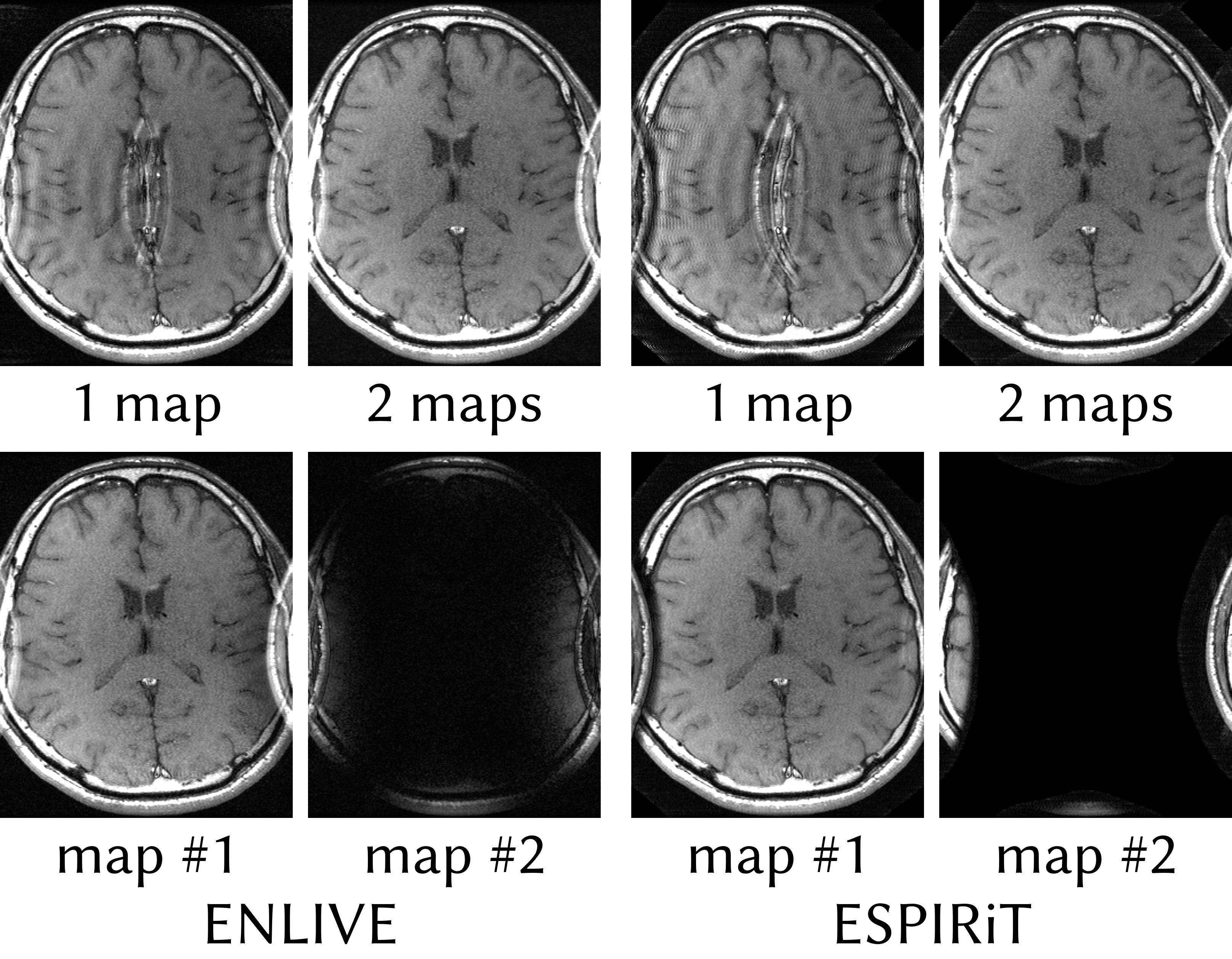}
	\caption{\label{fig:smallfov_brain}
	Comparison of ENLIVE and ESPIRiT
	reconstructions allowing both one and two sets of maps (top row)
	together with individual map images (bottom row) for the reconstructions
	using two maps.
	While the reconstructions using a single set of maps exhibit strong aliasing
	artifacts, the reconstructions allowing two sets of maps are artifact-free.
	The
	reason can be seen in the individual images: A single image
	with a single set of coil profiles cannot
	resolve the aliasing arising from the infolded sides. Using two sets of
	maps,
	the region causing infolding can be separated into the second image.
	}
\end{figure}

\begin{figure}[htbp]
\centering
\includegraphics[scale=\figscale]{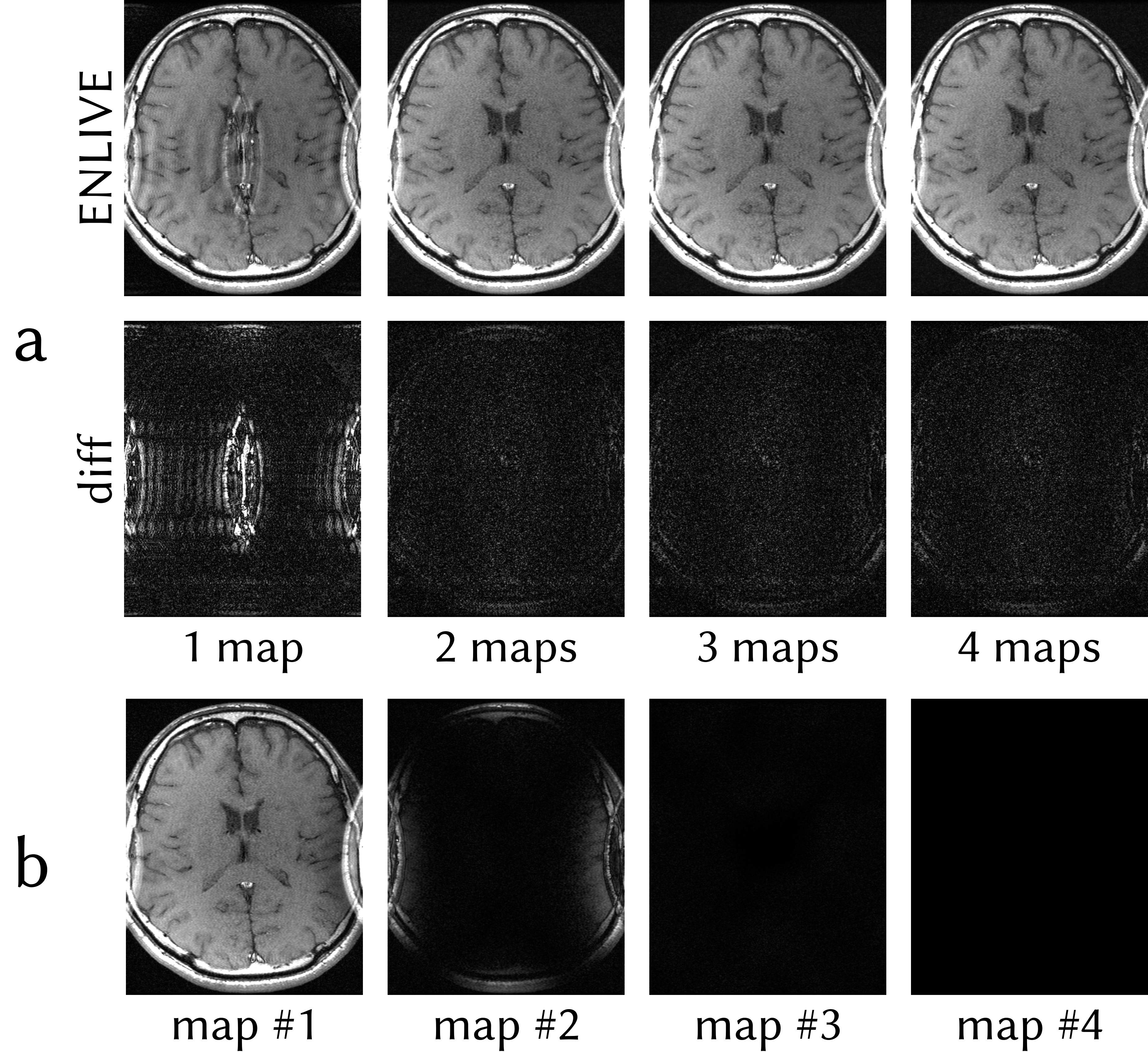}
	\caption{\label{fig:smallfov_maps}
	(a): ENLIVE reconstructions of the same data as in
	Fig. \ref{fig:smallfov_brain}
	using 1, 2, 3 and 4 sets of maps. Difference images to fully-sampled 
	reference data are shown in the bottom row. Using a single map, the 
	central artifact is clearly visible in the reconstruction as well as in the
	difference image. Using 2 and more maps, the artifact is resolved and 
	the difference images show close to no variation.
	(b): Individual map images of the reconstruction using 4 maps. Since 2 sets
	of maps are sufficient to fully describe the data, the first two maps are
	similar to the maps depicted in Fig. \ref{fig:smallfov_brain} while maps
	3 and 4 are close to zero.
	The corresponding coils profiles are depicted in
	Fig. \ref{fig:smallfov_sens}.
	}
\end{figure}

\begin{figure}[htbp]
	\centering
	\includegraphics[scale=\figscale]{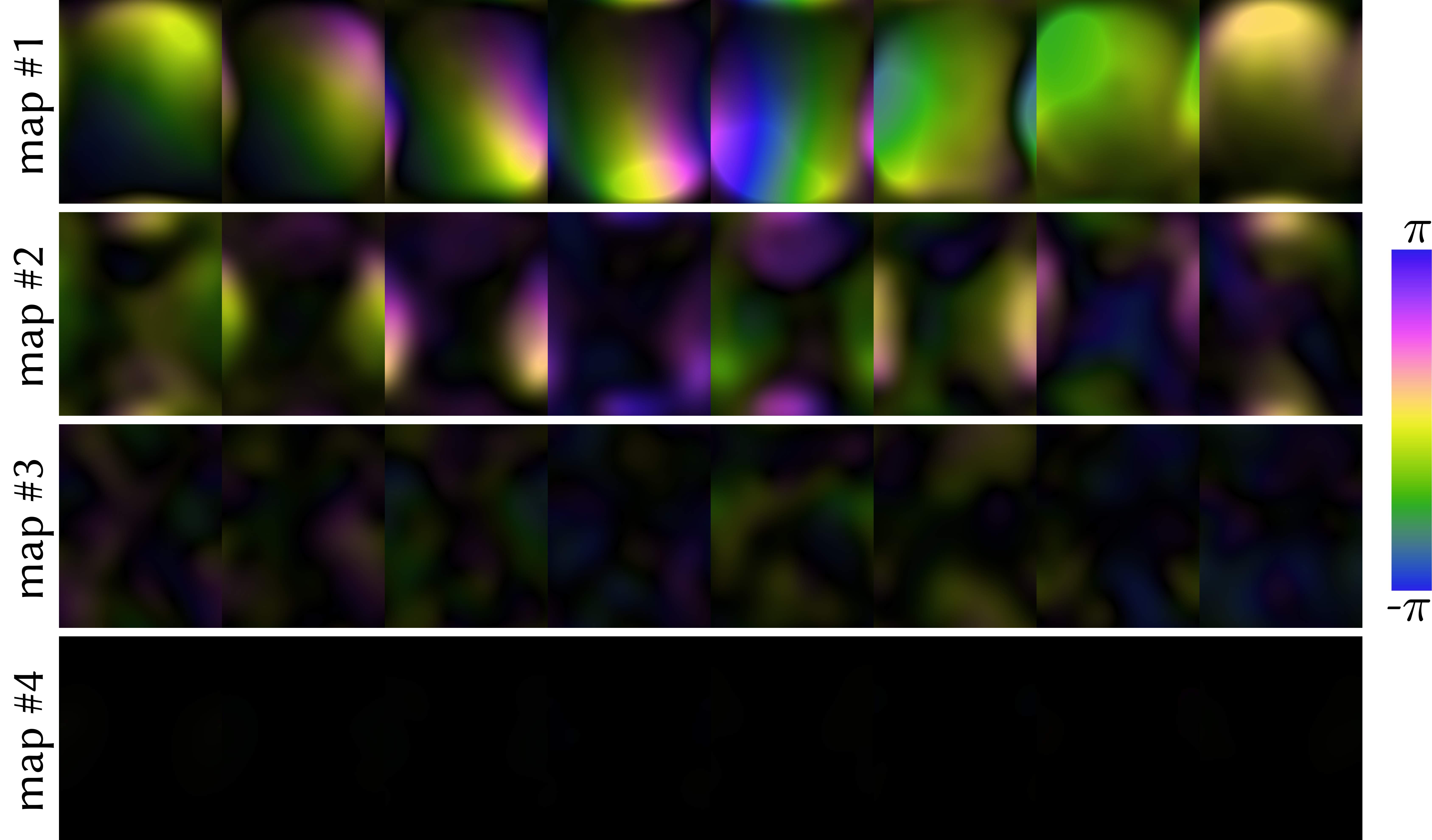}
	\caption{\label{fig:smallfov_sens}
		Calculated coil sensitivity profiles for the ENLIVE reconstruction 
		using 4 sets of maps shown in
		Fig. \ref{fig:smallfov_maps}.
		The second map is sensitive in the region
		which causes infolding in the single-map reconstruction, while the
		first map is smoothly sensitive over the entire FOV. The third
		and fourth map show very little sensitivity anywhere. Magnitude is 
		encoded in brightness while phase in encoded in the color, according to 
		the cyclic magenta-yellow-green-blue colormap described
		in~\cite{Kovesi_arXiv_2015}.
	}
\end{figure}

\begin{figure}[htbp]
\centering
\includegraphics[scale=\figscale]{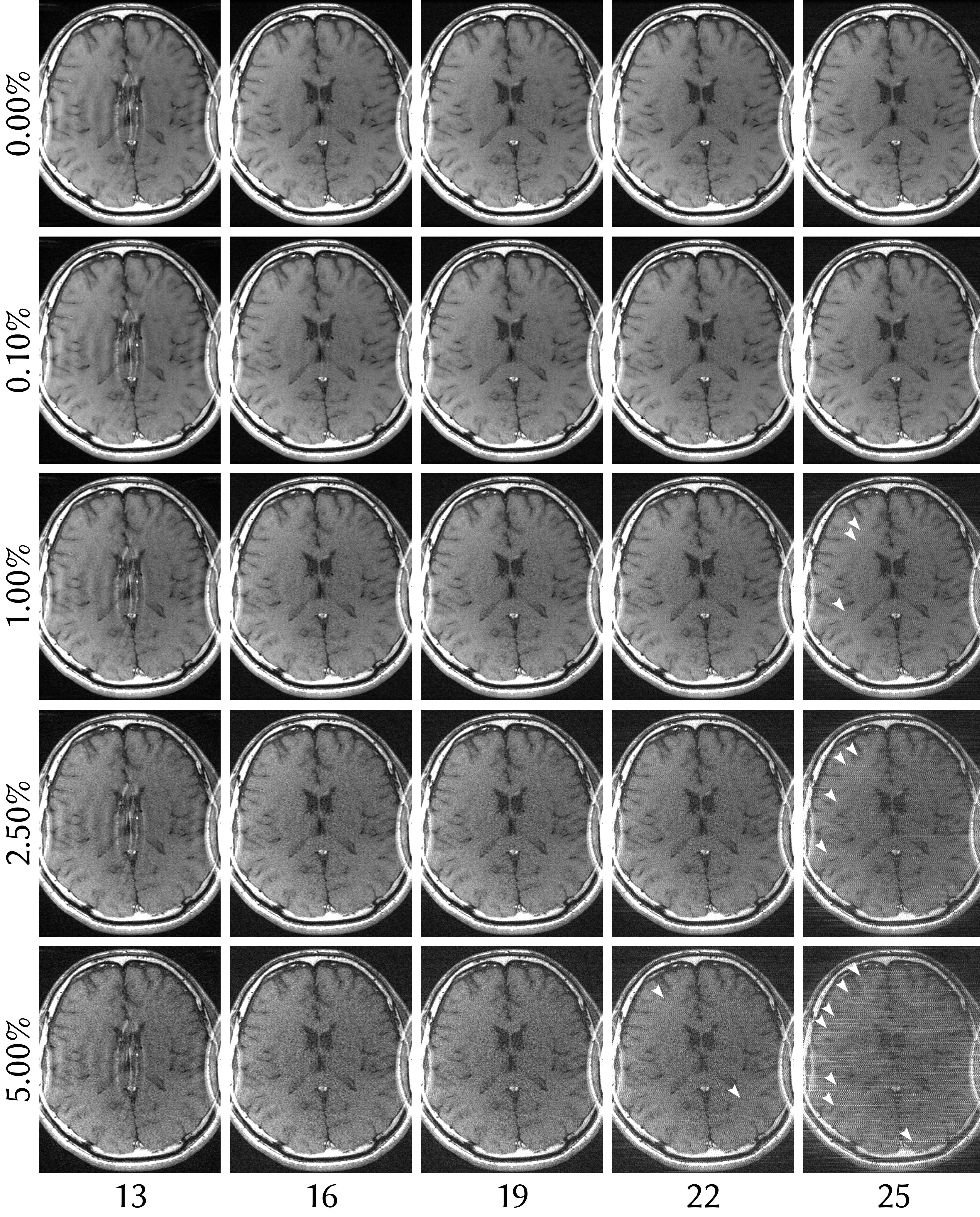}
\caption{\label{fig:newton_noise}
	ENLIVE reconstruction with 2 maps with differing number of Newtons steps
	(left to right) and different levels of added noise (top to bottom) of the
	same dataset as in \cref{fig:smallfov_brain}. Gaussian white noise was
	added to the k-space before reconstruction. The standard deviation of the
	added noise was varied between 0 and \SI{5}{\percent} of the absolute value
	of the DC component.
	Using too few Newton steps leads to residual infolding artifacts, while too
	many Newton steps cause high-frequency artifacts to appear (some of which 
	are indicated by arrows). Since the
	number of Newton steps controls the regularization in the IRGNM, we can
	understand these two effects as too much and too little regularization.
	In all cases, the added
	noise has no impact on the infolding artifact.
}
\end{figure}

\begin{figure}[htbp]
\centering
\includegraphics[scale=\figscale]{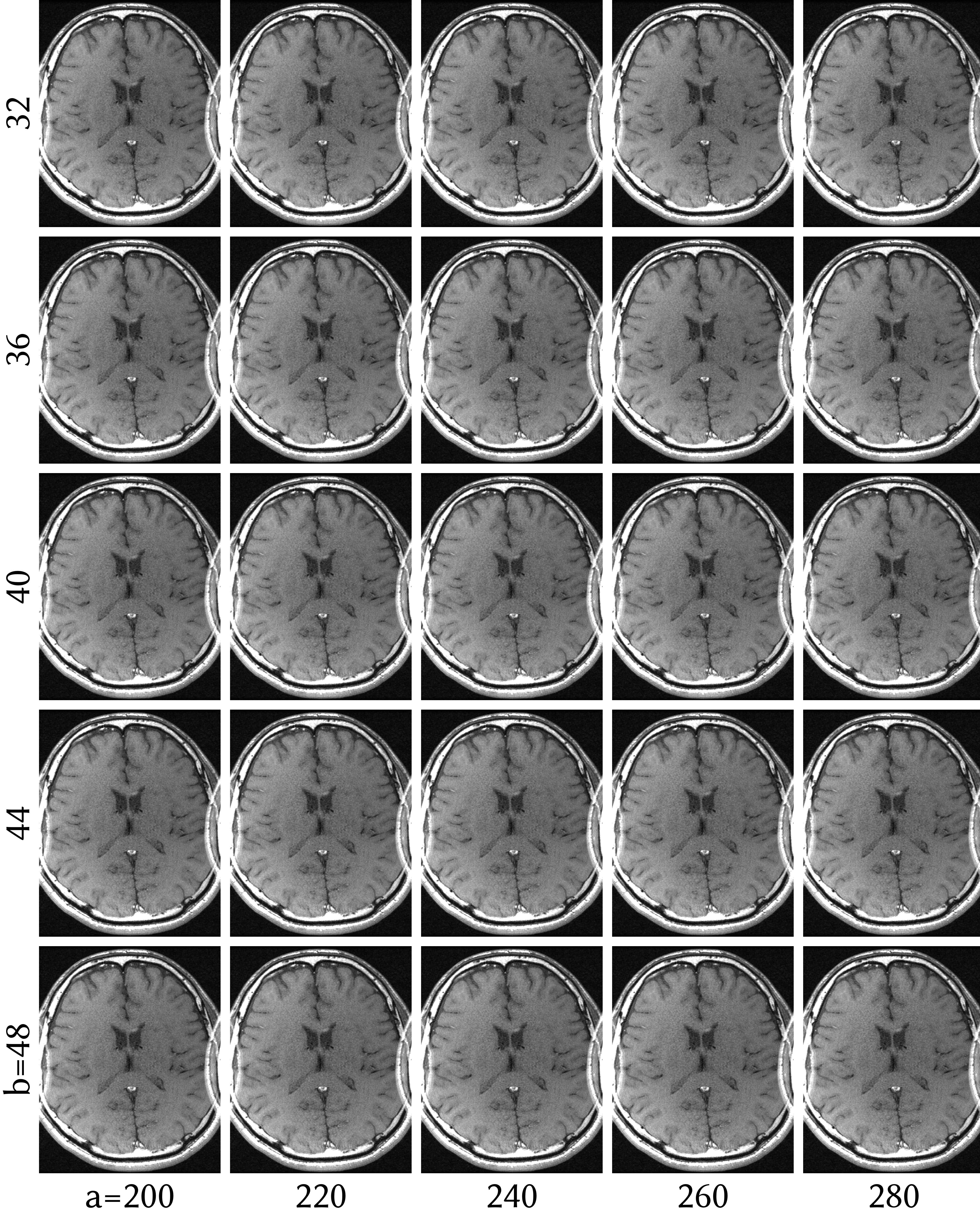}
\caption{\label{fig:a_b}
	ENLIVE reconstruction with 2 maps of the same dataset as in
	\cref{fig:smallfov_brain} with different parameters for the coil
	weighting matrix $\vect{W}$. $\vect{W}$ applies a penalty in k-space
	according to
	$(1+a\|\vect{k}\|^2)^{b/2}$. $a$ varies from left to right while $b$
	varies from
	top to bottom. For all other reconstructions, $a=240$ and $b=40$ (center
	image) were used. The infolding artifact does not appear for any parameter
	pair, indicating that the reconstruction is not sensitive to specific
	values of $a$ or $b$.
}
\end{figure}

\begin{figure}[htbp]
	\centering
	\includegraphics[scale=\figscale]{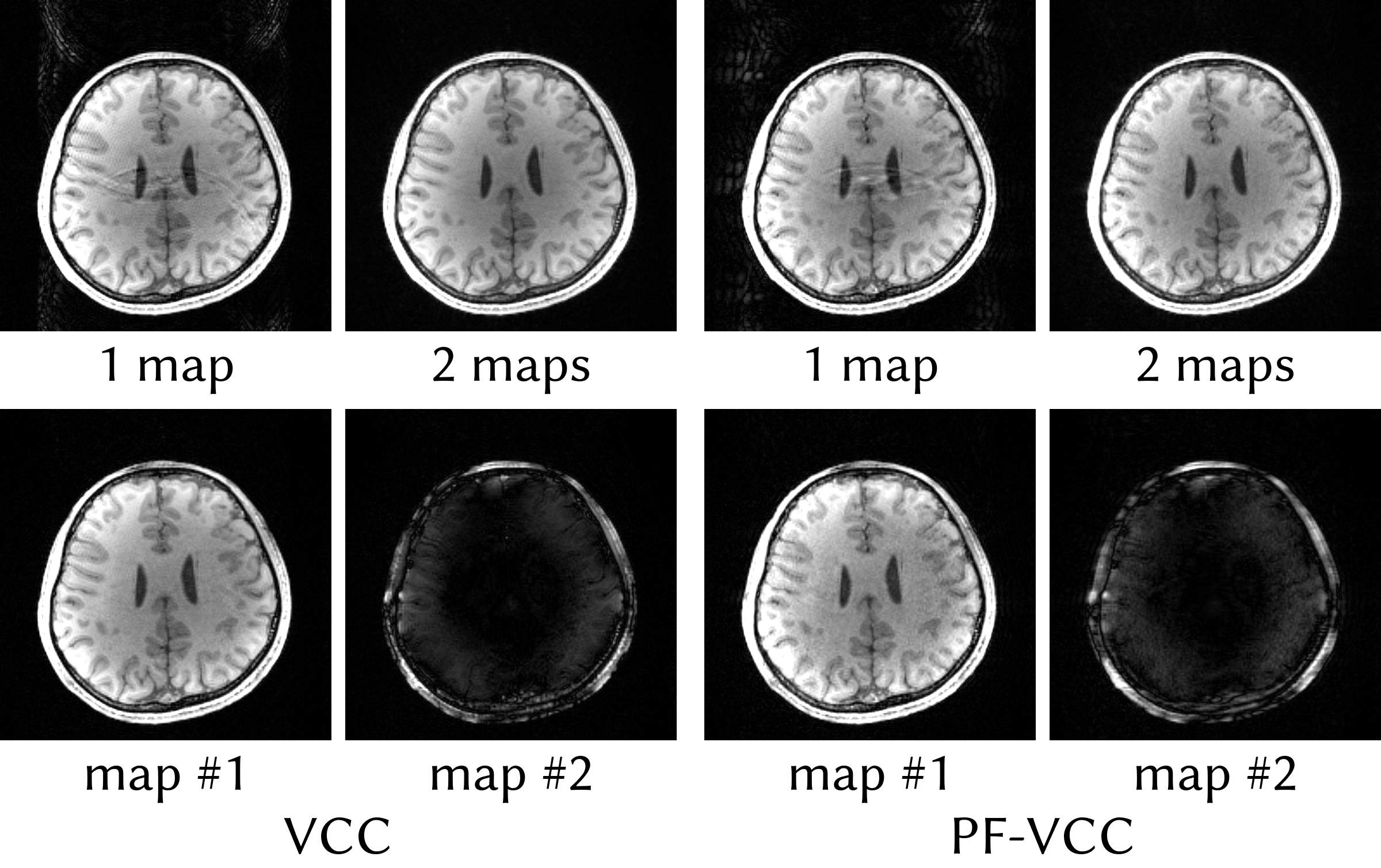}
	\caption{\label{fig:vcc} ENLIVE reconstructions allowing one and two sets of maps
		of data extended with virtual conjugate coils (VCC) and such data with
		a partial k-space (PF-VCC). The virtual-conjugate coils 
		impose a real-value constraint onto the data. High-frequency phase close
		to the skull violates this constraint, leading to artifacts in
		reconstructions using a single set of maps. By allowing two sets of maps,
		these regions with
		high-frequency phase variation are separated into the second image, 
		allowing almost artifact-free reconstruction. 
	}
\end{figure}

\begin{figure}[htbp]
	\centering
	\includegraphics[scale=\figscale]{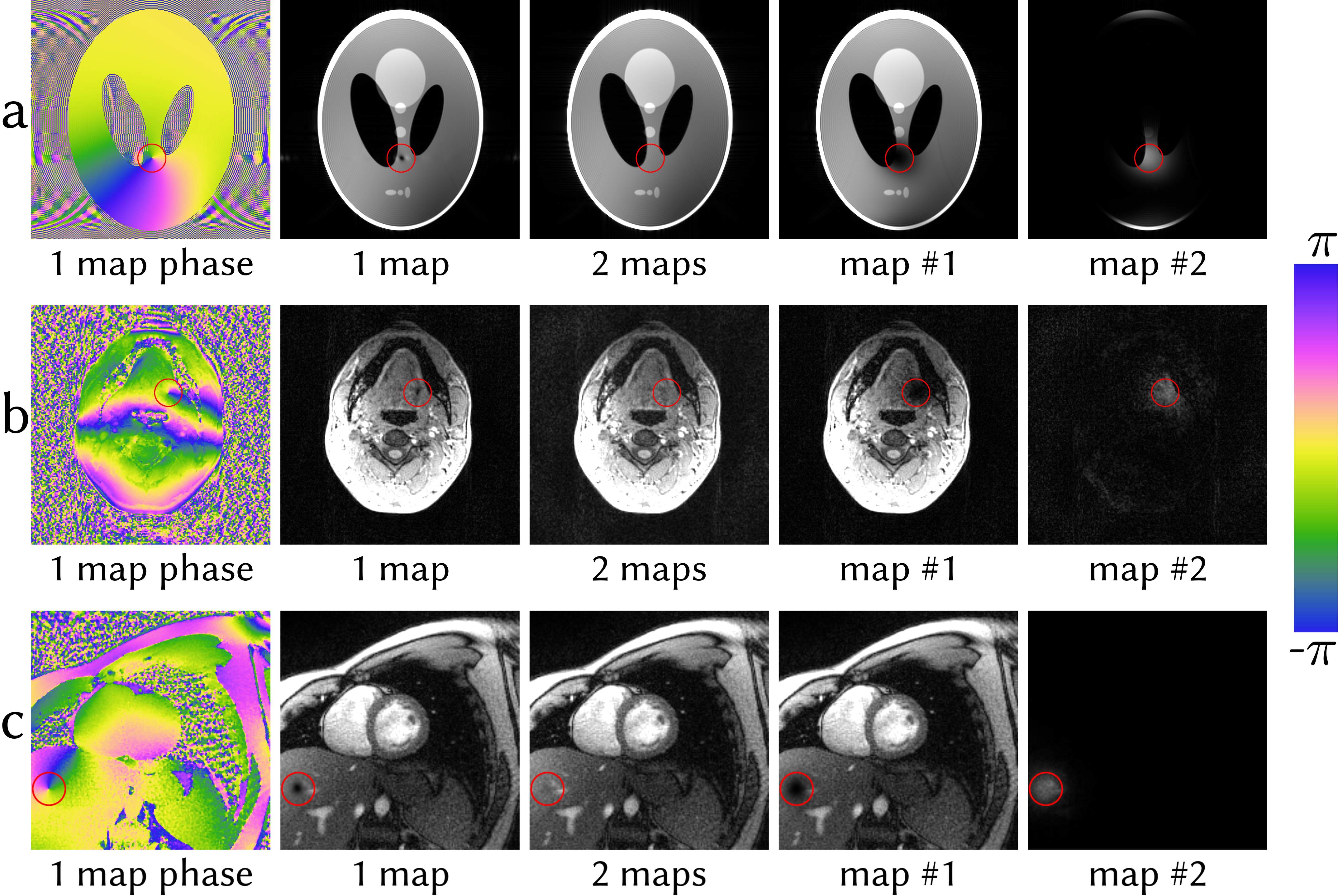}
	\caption{\label{fig:blackholes}
		Phase singularities in (a) a numerical phantom, (b) a transversal
		slice through the lower jaw and (c) a non-Cartesian short axis-view of 
		the human heart. Each dataset has been reconstructed
		with ENLIVE allowing one and two sets of maps. The phase singularity in
		(a) was produced by providing an initial guess containing a singularity.
		This singularity,
		clearly visible in the phase image, leads to artifactual signal loss at
		the same position in the post-processed magnitude image. 
		As in (a), the phase singularities in (b) and (c) lead to signal loss at
		the corresponding positions in the magnitude images.
		By allowing two sets of maps, ENLIVE
		can resolve this artifact by using the second set of sensitivities
		around the phase singularity, thereby providing an artifact-free 
		combined image. 
	}
\end{figure}

\begin{figure}[htbp]
\centering
\includegraphics[scale=\figscale]{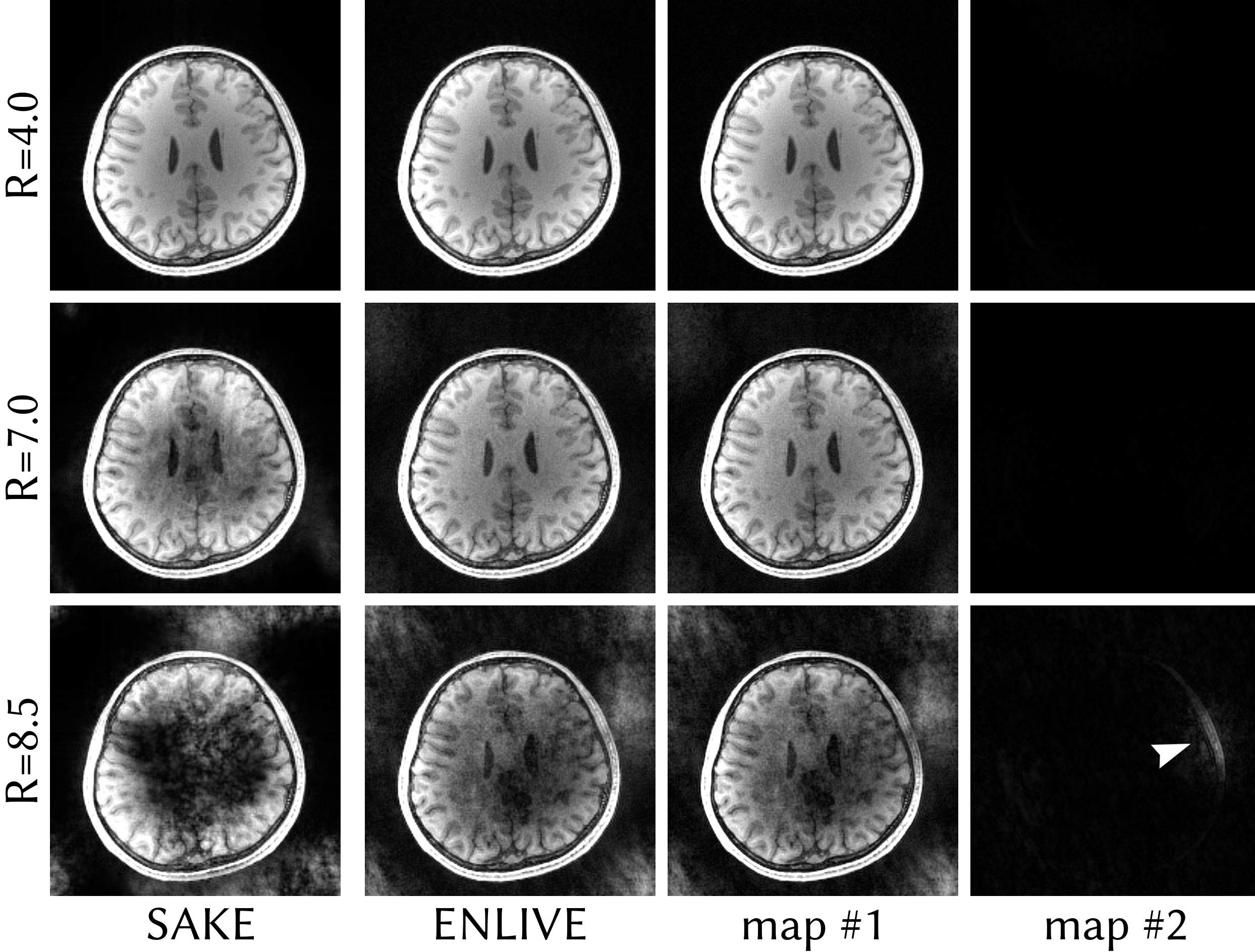}
\caption{\label{fig:empty}
		Variable-density Poisson-disc undersampled data with varying
		undersampling factors reconstructed with 
		ENLIVE allowing two sets of maps and with SAKE. The same slice as in
		Fig. \ref{fig:vcc} is used. Since this is a calibrationless parallel 
		imaging reconstruction without
		additional constraints and without model violations, a single set of
		maps is sufficient. For undersampling factors up to R=\num{7.0},
		ENLIVE therefore leaves the second allowed set
		empty, which causes the combined image to be essentially identical
		to the first set image. For an undersampling factor of R=\num{8.5},
		the ENLIVE reconstruction becomes very noisy and some image features
		start appearing the second map (indicated by an 
		arrow).
		For R=\num{4.0}, SAKE, too, provides artifact-free reconstruction. With
		higher undersampling factors artifacts appear in the images.
}
\end{figure}

\begin{figure}[htbp]
	\centering
	\includegraphics[scale=\figscale]{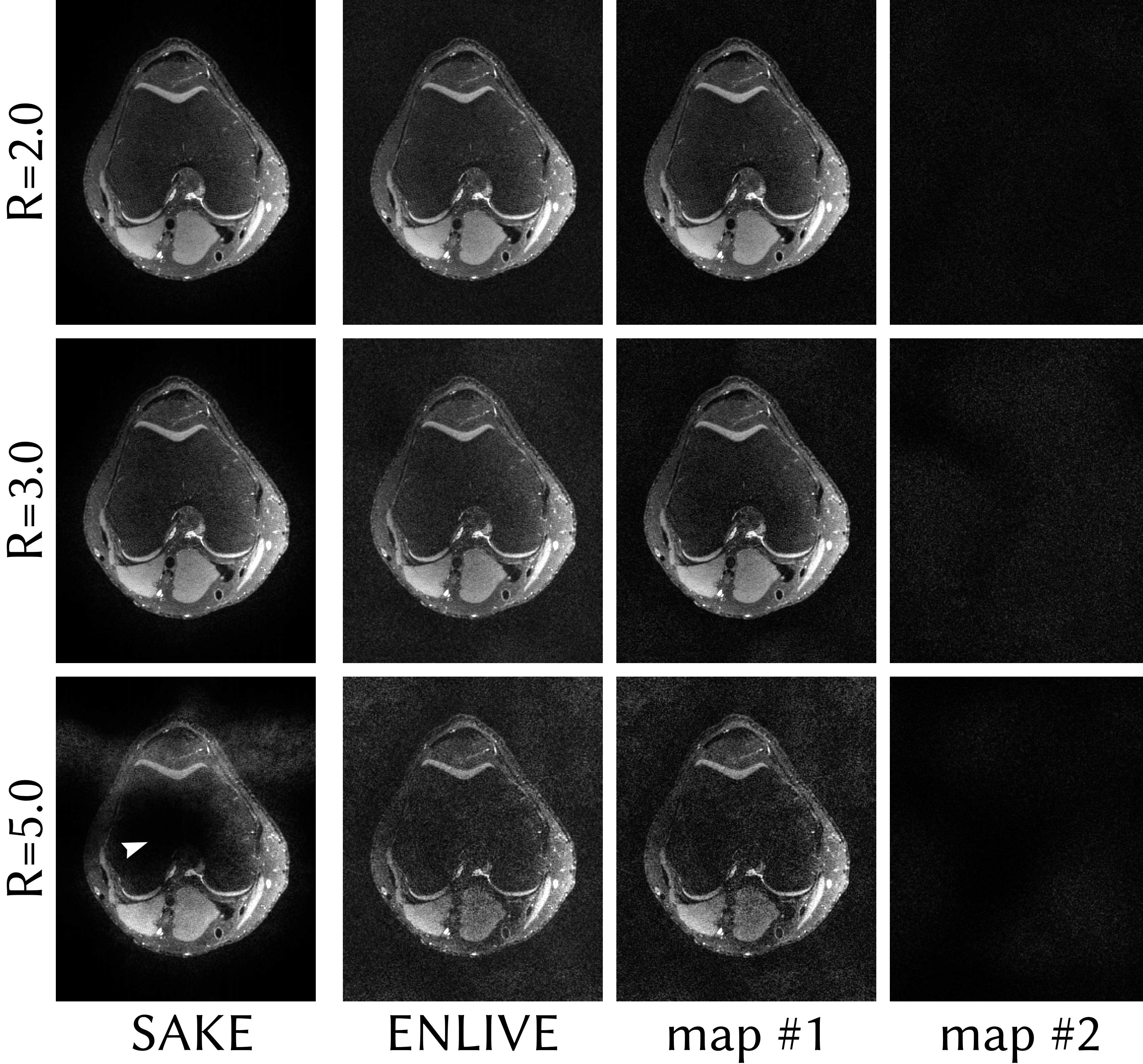}
	\caption{\label{fig:knee}
		Variable-density Poisson-disc undersampled data of a human
		knee with varying
		undersampling factors reconstructed with 
		ENLIVE allowing two sets of maps and with SAKE. This, too, is a 
		dataset without model violations. The second ENLIVE is therefore close 
		to zero. Up to R=\num{3.0}, both SAKE and ENLIVE provide artifact free
		reconstruction. For R=\num{5.0}, ENLIVE provides a reconstruction with
		high noise. SAKE, however, produces a large signal void in the 
		image center (indicated by an arrow).
	}
\end{figure}

\begin{figure}[htbp]
\centering
\includegraphics[scale=\figscale]{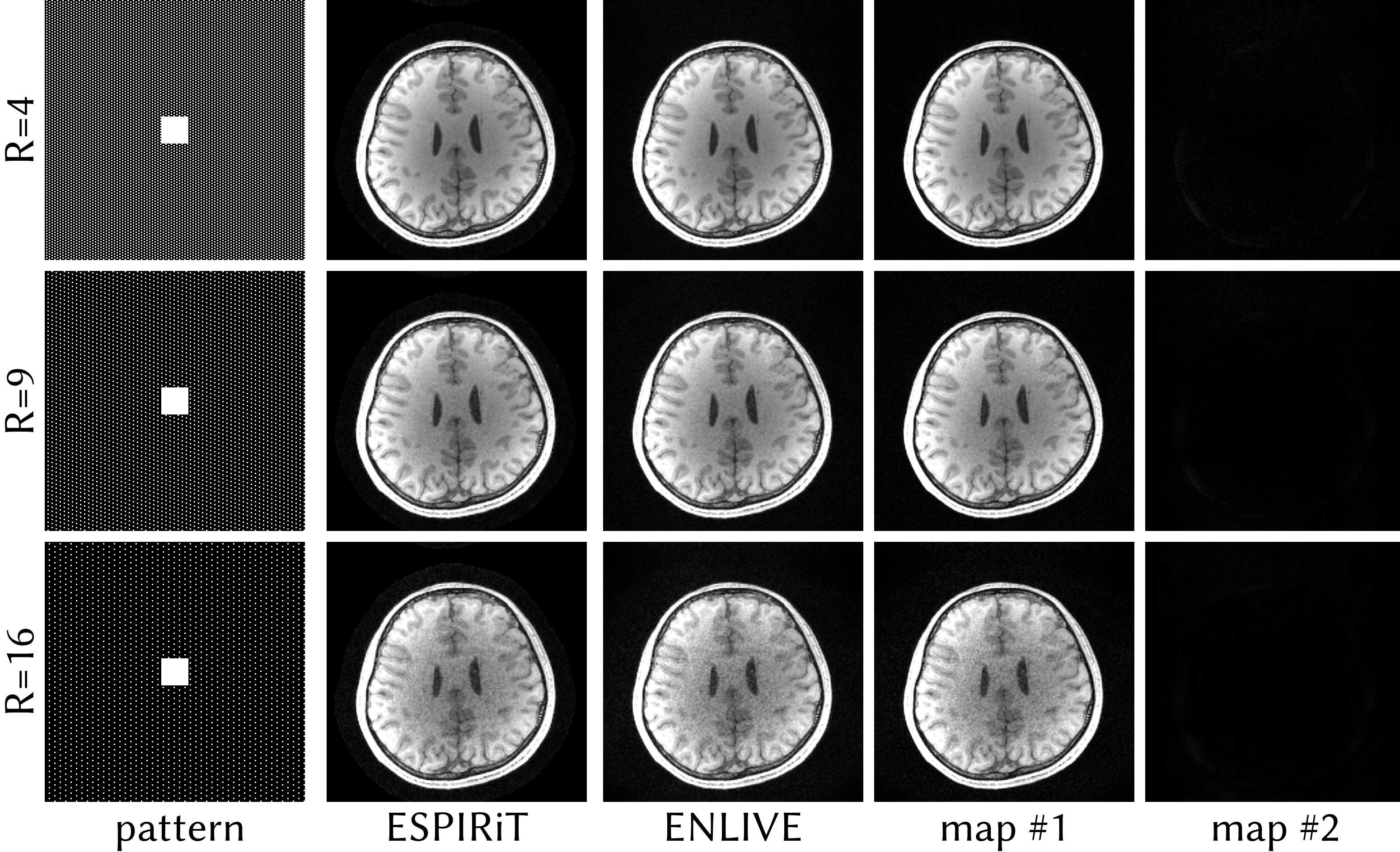}
\caption{\label{fig:high_unders}
	Comparison of ENLIVE reconstruction using 1 and 2 maps and ESPIRiT 
	reconstruction using 2 maps of the same dataset as in
	\cref{fig:vcc} undersampled with Cartesian CAIPIRINHA patterns
	with differing undersampling factors.
	Using two maps, ENLIVE and ESPRiT reconstructions show comparable quality. 
	Even though noise is increasing with higher undersampling, the second map
	remains close to zero. We conjecture that the adequate calibration region
	inhibits undersampling artifacts and ensures that no signal appears in the 
	second map, in contrast to \cref{fig:empty}.
}
\end{figure}

\subsection*{Limited FOV}
In the examples with a restricted FOV, both ENLIVE with a single set
of maps, i.e. NLINV, and ESPIRiT reconstructions show a similar central
artifact (\cref{fig:smallfov_brain}).
This artifact can be
readily explained as a consequence of the undersampling pattern and the signal
model violation at the edges of the image: Without a parallel imaging reconstruction,
we expect aliasing artifacts from all pixels in the FOV. The parallel
imaging reconstruction using a single set of maps
can resolve this aliasing only for pixels outside of the
regions of model violation. Since these edge regions alias to the image center,
the artifact appears there.
Both ENLIVE and ESPIRiT
reconstructions allowing multiple
sets of maps (\cref{fig:smallfov_brain,fig:smallfov_maps}a)
can resolve the aliasing everywhere.
For ENLIVE, the coil profiles
(\cref{fig:smallfov_sens}) of the second map are sensitive in these regions.
For ENLIVE using more than 2 sets of maps, the third and fourth map are
close to zero (\cref{fig:smallfov_maps}b).
Since no thresholding is used, they
cannot be exaclty zero. As is common in parallel imaging, tuning of the
regularization is necessary for successful reconstruction:
\cref{fig:newton_noise} shows that using too high
regularization (too few Newton steps) does not eliminate the central infolding
artifact, while too low regularization (too many Newton steps) leads to
high-frequency artifacts. Added noise degrades image quality, especially in the
case of too low regularization, but does not change the appearance of the
infolding artifact.
Additionally, \cref{fig:a_b} shows that the reconstruction is not sensitive to
specific choices for the parameters $a$ and $b$ of the coil weighting matrix
$W$.

\subsection*{Phase-constrained Imaging}
Next, reconstructions for phase-constrained imaging
using virtual-conjugate coils with and without an 
additional partial-Fourier factor are shown in \cref{fig:vcc}. In both cases,
reconstruction using only a single set of maps exhibit aliasing artifacts. These are
a consequence of the real-value constraint imposed by using virtual-conjugate
coils together with high-frequency phase variations caused by off-resonance from
fat: A single real-valued image cannot account for this high-frequency phase,
therefore the aliasing cannot be resolved.
Relaxing the reconstruction by allowing multiple sets of maps resolves this problem,
since the second set of maps can now account for this high-frequency phase variation.

\subsection*{Phase Singularities}
\cref{fig:blackholes}a shows a phantom example where the initial guess has been
intentionally chosen to induce a phase singularity in the reconstruction.
The phase singularity
leads to signal loss using a single set of maps. Using
ENLIVE allowing multiple sets of maps, the affected region can be resolved in
the second map. By combining the images, a single image without signal
loss can be recovered. This situation can also occur in practice.
\cref{fig:blackholes}b shows a slice through the throat with large phase
variations, while \cref{fig:blackholes}c shows a short-axis view of the human 
heart acquired with radial FLASH.
Using ENLIVE allowing multiple sets of maps, it is possible
to reconstruct artifact-free images.

\subsection*{Low-rank Property}
\Cref{fig:empty} and \cref{fig:knee} show calibrationless variable-density 
Poisson-disc
undersampled reconstructions with differing undersampling factors
comparing ENLIVE to SAKE.
In \Cref{fig:empty}, both ENLIVE
and SAKE provide
artifact-free reconstruction for moderate undersampling up to $R=4.0$.
At $R=7.0$, SAKE shows artifacts while
ENLIVE is artifact free. For these undersampling factors, the second ENLIVE set
image is close to zero, while the first set contains the image. For $R=8.5$,
both ENLIVE and SAKE show strong artifacts. Additionally, the second ENLIVE
map shows some image features.
Reconstruction time for $R=4.0$ for this dataset using a single core of an 
Intel Core i5-4590
CPU was \SI{22}{\second} using ENLIVE and \SI{6,3}{\hour} using SAKE.
In \Cref{fig:knee}, ENLIVE and SAKE provide artifact-free reconstruction up 
to $R=3.0$. At $R=5.0$, ENLIVE reconstruction is noisy while SAKE shows  a large
signal void. Reconstruction time for $R=2.0$ for this dataset using a single
core was\SI{18.6}{\second} using ENLIVE and \SI{41.5}{\minute} using SAKE.

\Cref{fig:high_unders} shows Cartesian ENLIVE reconstructions of data 
undersampled using CAIPIRINHA patterns with different undersampling factors. 
As a reference, the corresponding patterns are shown in the first column. For
all undersampling factors, the second map image is close to zero wile the first 
map contains the entire image. With increasing undersampling, high noise starts
to appear in the first map and the combned image. Still, no undersampling
artifacts appear even at $R=16$. Furthermore, even at this high undersampling,
no image features appear in the second map, in contrast to the result in
\cref{fig:empty}. We conjecture that the adequate calibration region in this
datasets prevents that artifact.

\section*{Discussion}

This work introduces ENLIVE, a nonlinear reconstruction method for parallel
imaging using a relaxed forward model. Using the IRGNM, 
ENLIVE simultaneously estimates
multiple sets of images and coil sensitivity profiles, extending NLINV 
by ESPIRiT's approach of using multiple sets of maps. The resulting bi-linear
problem with $\ell_2$-regularization can be related
to a lifted linear formulation using nuclear norm regularization, which
promotes low-rank solutions. From this, it becomes apparent that the method,
while employing a different parametrization,
is similar to SAKE and
P-LORAKS~\cite{Haldar_IEEETrans.Med.Imaging_2014,Haldar_Magn.Reson.Med._2016},
which are based on structured low-rank matrix completion
in k-space, and to CLEAR~\cite{Trzasko__2011}, which locally promotes
low-rankness in the image domain.
Although the low-rankness of the matrix considered in
the k-space methods is also caused by the fact that the signal lives
in a sub-space spanned by the coil sensitivities~\cite{Shin_Magn.Reson.Med._2014,Uecker_Magn.Reson.Med._2014},
it is constructed from many shifted copies of the signal in k-space.
This leads to a huge linear reconstruction problem with a rank constraint.
In contrast, CLEAR uses block-wise reconstruction in the image domain, which
is more similar to ENLIVE, but still requires a large number of small
SVDs.
A similar concept has been used to implement other low-rank methods.
For example, building on top of the work on object modeling introduced
in~\cite{Haacke_IEEETrans.Acoust.SpeechSignalProcessing_1989}, several
approaches using annihilating filters have recently been
proposed for combining parallel imaging with compressed
sensing~\cite{Jin_IEEETrans.Comput.Imag._2016,Lee_Magn.Reson.Med._2016,
	Ongie_IEEETrans.Comput.Imag._2017,Ongie_SIAMJ.Imag.Sci._2016}.
The existence of annihilating filters implies in turn the existence of a
weighted low-rank Hankel matrix which can be constructed from the k-space
samples. These methods then recover missing samples by structured low-rank
matrix completion.
In ENLIVE, the convex matrix completion problem has been replaced by
a much smaller bi-linear problem with simple quadratic
penalties~\cite{Davenport_IeeeJ.Sel.Top.Signa._2016,Recht_SIAMReview_2010}. In
some sense, this is similar
to the idea of transforming linear problems with $\ell_1$-regularization
into quadratic problems with $\ell_2$-regularization~\cite{Uecker__2008}.

Low-rank approaches have also been proposed for dynamic imaging.
One method for blind compressed
sensing~\cite{Lingala_IEEETrans.Med.Imag._2013} estimates both
the time series of images as well as a dictionary which sparsifies that series.
Haldar and Liang~\cite{Haldar__2010} introduce a method which uses partial
separability of the signal into functions describing its k-space and its time
dependence. Both of these approaches exploit the low rank of the time-dependent
signal. While structurally similar,Haldar and Liang~\cite{Haldar__2010} use
an explicitly 
rank-constraint formulation while
Lingala et al.~\cite{Lingala_IEEETrans.Med.Imag._2013} use
an $\ell_1$-norm to induce sparsity. In contrast, ENLIVE's
$\ell_2$-regularization achieves low-rankness without
an explicit rank constraint
through the equivalence to a formulation with regularization of the nuclear
norm outlined in the Theory, which forms the core of the proposed method.

ENLIVE can also be related to a previous extension of NLINV proposed
for separation of chemical
species\cite{Uecker__2012a,Shin_Magn.Reson.Med._2015}.
This method is based on the idea that the signal is a superposition of different
images shifted in the spatial domain according to the chemical shift.
As also shown for ESPIRiT, the sensitivities for the shifted signals
from different species also appear to be shifted. They therefore violate
the simple SENSE model with a single set of of maps and, consequently,
cause the appearance of a second set of maps. The previously proposed
extension to NLINV can be understood as a version of ENLIVE with the additional
constraint that different sets of sensitivities are shifted versions
of each other.

As shown in this work,
small FOV and phase-constrained
reconstructions using a single set of maps show
artifacts whenever there are inconsistencies which cannot be explained using
the simple model, while ENLIVE allowing two sets of maps enables
artifact-free reconstruction in all evaluated cases. When using correct 
regularization, added noise does not impede artifact removal either.
In cases where reconstruction
with a single set of maps is already free from artifacts, ENLIVE
automatically only uses a single set.
In general, though, the maximum number of ENLIVE
maps must be specified manually.
This is similar to ESPIRiT where, while theoretically the correct number of maps
can automatically be estimated as the multiplicity of the eigenvalue 1, in practice
a maximum number of maps is set in advance to enable efficient computation of the
eigenvector maps by power iteration. However, an extension to ENLIVE
to automatically adapt the number of maps during the iteration is also
conceivable.

As the distribution of the phase between image and coil sensitivities
cannot be determined from the data alone without additional prior
knowledge, choosing a good phase is a common problem when
calibrating sensitivities.\cite{Inati__2013,Bilgic__2016}
This fundamental problem affects different algorithms in different ways.
In Walsh's method~\cite{Walsh_Magn.Reson.Med._2000} or ESPIRiT a
pixel-wise phase across channels simply
remains undefined and has to be aligned to a reference.
If the reference is not ideal, phase singularities may occur.
Phase singularities imply a non-smooth phase which then reduces
sparsity in compressed sensing, preventing an efficient and compact
representation of the sensitivities in the Fourier domain~\cite{Uecker__2013a}, or
causing problems in post-processing.
For example, as Li et al.~\cite{Li_J.Magn.Reson.Imaging_2015}
have shown, phase singularities can appear as artifactual microhemorrhage
in susceptibility weighted imaging. NLINV and ENLIVE guarantee smooth
sensitivities, but this then
traps the algorithm in a local minimum and
creates a hole
instead.\cite{Wang_Magn.Reson.Med._2018}
For ENLIVE, the
use of a second set of maps may still avoid signal loss in the
reconstruction.

Even though local minima are a general concern with nonlinear methods, in
our experience, the only practically relevant examples are the phase
singularities. There, although the ENLIVE reconstruction is not optimal, use 
of a second map may mitigate the resulting artifact.

In summary, ENLIVE combines different advantages of NLINV, ESPIRiT, and SAKE.
As NLINV and SAKE, it utilizes all available data, can be directly
applied to non-Cartesian data, and does not require a calibration
region. As ESPIRiT and SAKE, it is not limited to the SENSE
model but automatically adapts to certain inconsistencies in
the data. As ESPIRiT and NLINV, it is computationally
efficient and makes use of an explicit image-domain
representation during reconstruction which facilitates
the use of advanced regularization terms.

\section*{Conclusion}

In this work we propose ENLIVE, a nonlinear method for
parallel imaging which seeks to combine the robustness of ESPIRiT with 
the flexibility of NLINV.
ENLIVE can be related to a lifted formulation of
blind multi-channel deconvolution with nuclear norm regularization,
which show that it belongs to the class of
 calibrationless parallel imaging methods
based on structured low-rank matrix completion.
In imaging settings involving
limited FOV, phase constraints, and phase singularities, it has been shown
to provide artifact-free reconstruction with quality comparable to
state-of-the-art methods.

\section*{Methods}

The proposed method was implemented in the Berkeley Advanced Reconstruction 
Toolbox (BART)~\cite{Uecker__2015} and all other reconstructions were
performed using BART as well.
Process-level par\-al\-leliza\-tion was
achieved 
using \texttt{GNU par\-al\-lel}.\cite{Tange_;login:TheUSENIXMagazine_2011}
To facilitate the reproducibility of our research,
data and source code used to generate the results of this paper can be 
downloaded
from \url{https://github.com/mrirecon/enlive}.

To test its robustness in case of inconsistencies, ENLIVE was applied in several
different experimental settings: We selected examples for imaging with an FOV 
smaller
than the extent of the object, phase-constrained imaging, and phase 
singularities.
In all cases, reconstructions using ENLIVE were performed using
one, i.e. NLINV, or two sets of maps
with initial regularization set to $\alpha_0=1$.
If not stated otherwise, 11 Newton steps
and $q=1/2$
were used for the IRGNM.
These parameters, as well as the parameters for the other
methods, were chosen according to best visual appearance.

In an example without inconsistencies
we tested whether ENLIVE produces results with only one set
of maps.
Additional examples show ENLIVE's performance under high undersampling
and in non-Cartesian imaging.

\subsection*{Limited FOV}

We applied ENLIVE to the same dataset used
in.\cite{Uecker_Magn.Reson.Med._2014}
This is a retrospectively 2-fold undersampled
2D spin-echo dataset (TR/TE~=~\SI{550/14}{\milli\second}, FA~=~\ang{90},
BW~=~\SI{19}{\kilo\hertz},
matrix size: \num{320x168},
slice thickness: \SI{3}{\milli\meter}, \num{24x24} calibration region) with an
FOV of
\SI{200 x 150}{\milli\meter}, acquired at
\SI{1.5}{\tesla} using an 8-channel head coil.
The dataset was zero-padded in k-space to produce square image space pixels.
This FOV is smaller than the head of the subject in the lateral
direction which leads to artifacts in a traditional
SENSE reconstruction.
These data were reconstructed with ENLIVE using one or two sets of maps
and compared to ESPIRiT using one or two sets of maps.
To investigate the effect of additional sets of maps, the data were
additionally reconstructed using 1, 2, 3, and 4 sets of maps.
For ENLIVE, $q=2/3$ was used. To investigate the sensitivity to noise and
to regularization, an additional reconstruction using \numlist{13;16;19;22;25}
Newton steps and added Gaussian white noise with noise levels of
\SIlist{0;0.1;1;2.5;5}{\percent} was performed. The noise level here is the
standard deviation of the added noise as percent of the magnitude of the
DC component. From this, 19 Newton steps was determined as the optimum and
used for reconstruction. For ESPIRiT a kernel size
of \num{6x6} and a threshold of \num{0.001} was used.

\subsection*{Phase-constrained Imaging}

Phase-constrained parallel imaging~\cite{Willig_J.Magn.Reson._2005} with 
virtual conjugate coils~\cite{Blaimer_Magn.Reson.Med._2009}
is equivalent to an explicit phase constraint in SENSE, but more robust in 
GRAPPA and
ESPIRiT due to their ability to adapt to 
inconsistencies~\cite{Blaimer_Magn.Reson.Med._2016,Uecker_Magn.Reson.Med._2017}.
To assess ENLIVE's performance in phase-constrained imaging settings with
virtual conjugate coils, we applied
it to the same dataset used in.\cite{Uecker_Magn.Reson.Med._2017}
This is a single slice in readout direction of a retrospectively
3-fold undersampled 3D FLASH dataset
(TR/TE~=~\SI{11/4.9}{\milli\second})
acquired at \SI{3}{\tesla} using a 32-channel head coil. \num{24x24}
auto-calibration lines were used. Additionally, a partial Fourier factor
of $5/8$ was applied to the data and evaluated separately.

\subsection*{Phase Singularities}

Similar to other
algorithms~\cite{Bilgic__2016,Inati__2013,Wang_Magn.Reson.Med._2018}
phase singularities can appear in coil sensitivity
profiles with ENLIVE. As ENLIVE enforces smooth coil
sensitivity profiles, this leads to an artifactual hole in the
sensitivities around the singularity.
To demonstrate this effect, we synthetically constructed an
example using BART to generate 6-channel k-space data
(matrix size: \num{256 x 256})
of the numerical
Shepp-Logan phantom. To get ENLIVE trapped in a local minimum
with a phase singularity, we provided an initial guess already containing
a phase singularity.
In regions with rapid phase variation, such phase singularities
can also appear in ENLIVE
reconstructions of in-vivo data. A transversal slice
through the throat containing such a phase singularity
was selected from the same dataset used for phase-constrained imaging.

To further show that ENLIVE can be applied directly to non-Cartesian data, we
reconstructed selected data containing a phase singularity from a real-time
FLASH~\cite{Uecker_Magn.Reson.Med._2010}
acquisition using a 30 channel thorax coil
of a short-axis view through the heart
of a volunteer with no known illnesses
(TR/TE~=~\SI{2.22/1.32}{\milli\second}, FA~=~\ang{10},
matrix size: \num{160x160}, FOV~=~\SI{256 x 256}{\milli\meter},
slice thickness: \SI{6}{\milli\meter}, field strength: \SI{3.0}{\tesla}).
Five consecutive frames during diastole, comprising 65 radial spokes,
were selected, corrected for gradient 
delays~\cite{Moussavi_Magn.Reson.Med._2013},
regridded to a $1.5$ times finer grid and subsequently
reconstructed with ENLIVE using 1 and 2 maps.
For this dataset, $q=2/3$ and 17 iterations of the IRGNM were used.

\subsection*{Low-rank Property}

In order to show that ENLIVE automatically uses only the required number of
sets of maps, we retrospectively undersampled the same 3D dataset used for 
phase-constrained
imaging using variable-density Poisson-disc
sampling~\cite{Vasanawala__2011}
with undersampling factors of $R=4.0,\, 7.0,\, 8.5$ and
without a calibration region, and then extracted the same slice in readout 
direction. As a comparison, these data were also
reconstructed using
SAKE with \num{50} iterations and a relative size of the signal
subspace of \num{0.05}.

Additionally, we applied SAKE and ENLIVE to a
3D fast spin-echo acquisition~\cite{Hennig_Magn.Reson.Med._1986}
of a human knee
(TR/TE~=~\SI{1550/25}{\milli\second}, FA~=~\ang{90},
echo train length:~\num{40},
matrix size: \num{320x256}, FOV~=~\SI{160 x 153.6}{\milli\meter}, field 
strength: \SI{3.0}{\tesla}) from 
\url{mridata.org}~\cite{Ong__2018}. 
This dataset
was also undersampled using variable-density Poisson-disc sampling with 
undersampling factors of $R=2,\, 3,\, 5$ and a single slice in readout direction
was extracted. These data were then reconstructed using ENLIVE with 1 and 2 
maps and with SAKE with \num{50} iterations and a relative size of the signal
subspace of \num{0.125}.

To evaluate ENLIVE in settings with high acceleration factors, we
undersampled the 3D dataset used for 
phase-constrained imaging using Cartesian 
CAIPIRINHA~\cite{Breuer_Magn.Reson.Med._2006}
patterns with undersampling factors of $R=4,\, 9,\, 16$ with a \num{24x24}
calibration region. These data were then reconstructed with ENLIVE using
2 maps with $q=1/3$ and $8$ iterations of the IRGNM.

\appendix
\section*{Appendix}
\newcommand\HL{\bm}
\subsection*{Operator $\mathcal{A}$}
Here, we show the layout of $\vect{u}$, $\vect{v}$, and $\vect{u}\vect{v}^T$ as 
well as the action of
$\mathcal{A}$ using an image of size $N_{I}\defd n_x\cdot n_y \cdot n_z$ and 
$N_{C}$ 
coils.
Then, the vector $\vect{u}\equiv \vect{m}\in\mathcal{C}^{N_{I}}$
is defined as
\begin{align}
\vect{u}^T = \begin{pmatrix}
m_1 &
\dots &
m_{N_{I}}
\end{pmatrix}
\end{align}
and the vector $\vect{v}\in\mathcal{C}^{N_{C}\cdot N_{I}}$ 
of stacked, weighted coil sensitivity profiles $\hat{\vect{c}}_j$ as
\begin{align}
\vect{v}^T = \begin{pmatrix}
\hat c_{1,1} &
\dots  &
\hat c_{N_{I},1} &
\hat c_{2,2} &
\dots &
\hat c_{N_{I},N_{C}}
\end{pmatrix}
\end{align}
where $\hat c_{ij}$ is the $i$th pixel of the $j$th weighted coil profile.
Therefore, $\vect{u}\vect{v}^T\in\mathcal{C}^{N_{I}\times N_{C}\cdot N_{I}}$ is
\begin{align}
\vect{u}\vect{v}^T =
\begin{pmatrix}
{m_1\hat c_{1,1}} & m_1\hat c_{2,1}  & \dots & m_1\hat c_{N_{I},1}
& {m_1\hat c_{1,2}} & m_1\hat c_{2,2} & \dots & m_1\hat c_{N_{I},N_{C}} \\
m_2\hat c_{1,1} & {m_2\hat c_{2,1}}  & \dots & m_2\hat c_{N_{I},1}
& m_2\hat c_{1,2} & {m_2\hat c_{2,2}} & \dots & m_2\hat c_{N_{I},N_{C}} \\
\vdots & \vdots & \ddots & \vdots & \vdots & \vdots & & \vdots \\
m_{N_{I}}\hat c_{1,1} & m_{N_{I}}\hat c_{2,1} & \dots & {
m_{N_{I}}\hat c_{N_{I},1}}
& m_{N_{I}}\hat c_{1,2} & m_{N_{I}}\hat c_{2,2} & \dots & {
m_{N_{I}}\hat c_{N_{I},N_{C}}}
\end{pmatrix}
\end{align}
Applying the inverse of the weighting matrix $\vect{W}$ yields
\begin{align}
\vect{u}\vect{v}^T\vect{W}^{-1} =
\begin{pmatrix}
\HL{m_1c_{1,1}} & m_1c_{2,1}  & \dots & m_1c_{N_{I},1}
& \HL{m_1c_{1,2}} & m_1c_{2,2} & \dots & m_1c_{N_{I},N_{C}} \\
m_2c_{1,1} & \HL{m_2c_{2,1}}  & \dots & m_2c_{N_{I},1}
& m_2c_{1,2} & \HL{m_2c_{2,2}} & \dots & m_2c_{N_{I},N_{C}} \\
\vdots & \vdots & \ddots & \vdots & \vdots & \vdots & & \vdots \\
m_{N_{I}}c_{1,1} & m_{N_{I}}c_{2,1} & \dots & \HL{
	m_{N_{I}}c_{N_{I},1}}
& m_{N_{I}}c_{1,2} & m_{N_{I}}c_{2,2} & \dots & \HL{
	m_{N_{I}}c_{N_{I},N_{C}}}
\end{pmatrix}
\end{align}
The diagonals containing products of image pixels with corresponding
coil profile pixels are highlighted in bold. The action of the operator
$\mathcal{A}$ is to select these highlighted entries of 
$\vect{u}\vect{v}^T\vect{W}^{-1}$, apply a
two or three dimensional Fourier transform to each coil image
and finally apply a  mask
$\mathcal{P}$ projecting
onto the acquired pattern.

\subsection*{Equivalence of formulations}
In the following we will show that
the lifted rank-$k$ problem in Eq. \eqref{eq:lifted_nlinv} corresponds to
the ENLIVE formulation in Eq. \eqref{eq:enlive}.
Using the linearity of
the operators, the following holds:
\begin{align}
\mathcal{A}\left\{\vect{U} \vect{V}^T\right\} & = \mathcal{A} \left\{
\sum_{i=1}^{k} \vect{u}_{i}
\vect{v}_{i}^T \right\} \\
& = \sum_{i= 1}^{k} \mathcal{A} \left\{
\vect{u}_{i} \vect{v}_{i}^T \right\}  \\
& = \sum_{i=1}^{k} \left( P \mathcal{F} \left\{
\vect{c}^{i}_j\odot \vect{m}^{i} \right\} \right)_{j=1\dots N_{c}}\\
& = \left(P \mathcal{F} \left\{ \sum_{i=1}^k
\vect{c}^{i}_j\odot \vect{m}^{i} \right\} \right)_{j=1\dots N_{c}}
\end{align}
Here, we make use of the definition of the operator $\mathcal{A}$:
\begin{align}
\mathcal{A} \left\{ \vect{u}_{i} \vect{v}_{i}^T \right\} :=
\left( P\mathcal{F} \left\{ \vect{c}^{i}_j\odot \vect{m}^{i}
\right\} \right)_{j=1 \dots N_{c}}
\end{align}

\bibliography{radiology}

\section*{Acknowledgements}

Supported by the DZHK (German Centre for Cardiovascular Research).
Part of this research was funded by the
Physics-to-Medicine Initiative Göttingen 
(LM der Niedersächsischen Vorab) and
DFG (UE 189/1-1). We acknowledge support by the Open Access Publication Funds 
of the Göttingen University.

\section*{Author contributions statement}
All authors contributed to the design of the study.
H.C.M.H., S.R., and M.U. implemented the method. H.C.M.H.
performed the numerical experiments. H.C.M.H., R.N.W.
and M.U. contributed to the data analysis.
F.O. and M.L. provided guidance on design and implementation.
All authors contributed to the preparation of
the manuscript.

\iftoggle{scirep}{
\section*{Additional information}

\textbf{Competing financial interests}: The authors declare no competing 
interests.

\newpage
\begin{NoHyper}
\paragraph{\Cref{fig:smallfov_brain}:}	\nameref{fig:smallfov_brain}
\paragraph{\Cref{fig:smallfov_maps}:}	\nameref{fig:smallfov_maps}
\paragraph{\Cref{fig:smallfov_sens}:}	\nameref{fig:smallfov_sens}
\paragraph{\Cref{fig:newton_noise}:}	\nameref{fig:newton_noise}
\paragraph{\Cref{fig:a_b}:}				\nameref{fig:a_b}
\paragraph{\Cref{fig:vcc}:}				\nameref{fig:vcc}
\paragraph{\Cref{fig:blackholes}:}		\nameref{fig:blackholes}
\paragraph{\Cref{fig:empty}:}			\nameref{fig:empty}
\paragraph{\Cref{fig:knee}:}			\nameref{fig:knee}
\paragraph{\Cref{fig:high_unders}:}		\nameref{fig:high_unders}
\end{NoHyper}

}

\end{document}